\documentclass[aps, prl, sd,  superscriptaddress, reprint, twocolumn]{revtex4-1}

\usepackage{amsmath}
\usepackage{graphicx}
\usepackage{dcolumn}
\usepackage{bm}
\usepackage{float}

\begin{document}

\title[]{
Nuclear Quantum Effects 
in Scattering of H and D from Graphene
}

\author{Hongyan Jiang*}
\affiliation{Department of Dynamics at Surfaces, Max-Planck-Institute for Biophysical Chemistry, \\ Am Faßberg 11, 37077 Göttingen, Germany}
\author{Xuecheng Tao*}
\affiliation{Division of Chemistry and Chemical Engineering, California Institute of Technology, \\ Pasadena, California 91125, USA}
\author{Marvin Kammler}
\affiliation{Department of Dynamics at Surfaces, Max-Planck-Institute for Biophysical Chemistry, \\ Am Faßberg 11, 37077 Göttingen, Germany}
\author{Feizhi Ding}
\affiliation{Division of Chemistry and Chemical Engineering, California Institute of Technology, \\ Pasadena, California 91125, USA}
\author{Alec M. Wodtke}
\affiliation{Department of Dynamics at Surfaces, Max-Planck-Institute for Biophysical Chemistry, \\ Am Faßberg 11, 37077 Göttingen, Germany}
\affiliation{Institute for Physical Chemistry, Georg-August University of Göttingen, \\ Tammanstraße 6, 37077 Göttingen, Germany}
\affiliation{International Center for Advanced Studies of Energy Conversion, Georg-August University of Göttingen, \\ Tammanstraße 6, 37077 Göttingen, Germany}
\author{Alexander Kandratsenka}
\affiliation{Department of Dynamics at Surfaces, Max-Planck-Institute for Biophysical Chemistry, \\ Am Faßberg 11, 37077 Göttingen, Germany}
\author{Thomas F. Miller III}
\affiliation{Division of Chemistry and Chemical Engineering, California Institute of Technology, \\ Pasadena, California 91125, USA}
\author{Oliver B\"{u}nermann}
\affiliation{Department of Dynamics at Surfaces, Max-Planck-Institute for Biophysical Chemistry, \\ Am Faßberg 11, 37077 Göttingen, Germany}
\affiliation{Institute for Physical Chemistry, Georg-August University of Göttingen, \\ Tammanstraße 6, 37077 Göttingen, Germany}
\affiliation{International Center for Advanced Studies of Energy Conversion, Georg-August University of Göttingen, \\ Tammanstraße 6, 37077 Göttingen, Germany}

\date{\today}

\begin{abstract}
We present a detailed study of the nuclear quantum effects in H/D sticking to graphene, comparing classical, quantum and mixed quantum/classical simulations to results of scattering experiments. Agreement with experimentally derived sticking probabilities is improved when nuclear quantum effects are included using ring polymer molecular dynamics. Specifically, the quantum motion of the carbon atoms enhances sticking, showing that an accurate description of graphene phonons is important to capturing the adsorption dynamics. We also find an inverse H/D isotope effect arising from Newtonian mechanics. 
\end{abstract}
\maketitle

\textit{Introduction. --}  
Nuclear quantum effects (NQEs) are ubiquitous in chemistry, typically resulting from quantum resonances, zero-point energy or tunneling. H/D substitution forms the basis of nearly all experiments designed to reveal NQEs, as the resulting energy shifts to the quantum levels influence differential reactive scattering cross-sections \cite{yang2008dynamical}, alter rate constants or even shift equilibria in chemical reactions \cite{kwart1982temperature}.

But NQEs in chemistry are far from being fully understood. For example, the {\it normal} isotope effect where H reacts faster than D is scarcely more common than the {\it inverse} isotope effect \cite{heazlewood2020strong,gomez2011kinetic,churchill2003normal}. Furthermore, heavy atom NQEs are also known \cite{laidler1987chemical}. Obviously, phonons in solids are quantum mechanical; yet often they are modelled with classical mechanics due to the solid's high dimensionality.
Little is known about the nature of errors introduced by a classical phonon approximation
in the context of interfacial chemical reactivity. 

Within this context, graphene is an ideal test system to examine NQEs in surface chemistry. The combination of light C-atoms and stiff C-C bonds means that graphene exhibits perhaps the highest frequency phonons of any common solid, meaning that the classical approximation of phonons could fail dramatically.

The simplest reaction on graphene is adsorption of an H or a D atom. A physisorption well with a depth of $\sim$40 meV is found at C-H distances of $\sim$4 \r{A} \cite{ghio1980vibrational,lepetit2011sticking} and at a C-H distance of $\sim$1.1 \r{A}, there is an $\sim$800 meV deep chemisorption well. Since chemisorption involves sp$^{2}$ to sp$^{3}$ rehybridization of a C atom, there is a pronounced barrier between the physisorbed and chemisorbed states \cite{sha2002first,zecho2002adsorption}. Saturated H/D atom adsorption shows an inverse isotope effect, possibly due to the interplay between the adsorption, reflection and associative desorption \cite{paris2013kinetic}. Hydrogen ions also penetrate through a graphene sheet faster than deuterium ions \cite{lozada2016sieving}, and it has been suggested that substantial barrier-lowering to hydrogen sticking on graphene occurs at low temperature \cite{ davidson2014cooperative}.

Recently, H atom scattering experiments were combined with first-principles theory to show that H collisions at graphene induce concerted in-plane motion of the carbon atom framework and extraordinarily fast energy dissipation leading to chemisorption \cite{jiang2019imaging}. The experiments produced nearly mono-energetic H atoms and could be performed at near-zero coverage, removing well-known ambiguities associated with the energy and coverage dependence of C-H bond formation on graphene \cite{hornekaer2006clustering,casolo2009understanding}.

In the current paper, we extend these scattering experiments to include both H and D, and we analyze the role of NQEs using a first-principles quantized molecular dynamics approach.
We developed a potential energy surface (PES) reparametrized to match electronic structure data obtained with a hybrid exchange-correlation functional, while maintaining a high level of numerical efficiency. This represents a significant improvement over our previously reported PES \cite{jiang2019imaging}. The new PES more accurately reproduces the graphene phonon density of states spectrum (PDOS), which proves to be essential for capturing the calculated NQE in this process.

Experiment provides no evidence of an H/D isotope effect; however, theory reveals a Newtonian isotope effect \cite{kelly2009newtonian,andujar2012racing} favoring D sticking over H that is similar in magnitude to the experimental error bars\textit. The longer interaction time of D with the graphene flake compared to H leads to increased sticking. Using Ring Polymer Molecular Dynamics (RPMD), we demonstrate mixed quantum/classical calculations where only selected degrees of freedom are treated quantum mechanically. This approach shows that the largest NQEs in this system are associated with the C-atom motion of the graphene\textit{---}these NQEs are not present when using a PES that fails to reproduce the high frequency region of graphene's PDOS spectrum. 

\textit{Experiment. --}
The experimental setup has been described in detail in Ref. \cite{bunermann2015electron,jiang2019inelastic,bunermann2018ultrahigh}.
All D experimental measurements are newly reported in this work.
H/D atoms are produced via photodissociation of a supersonic molecular beam of hydrogen/deuterium iodide with a 10 ns pulsed KrF laser producing atoms with incidence energy E$_i$ $\sim$ 1.0 eV. A portion of the atoms go through a skimmer and two differential pumping stages, enter the ultra-high vacuum chamber and scatter from the graphene sample. The sample is held on a 6-axis manipulator, allowing variation of the incidence angle $\vartheta_i$. About 0.7 mm above the surface, the scattered H or D atoms are excited to a long-lived Rydberg state ($n=34$) by two spatially and temporally overlapped laser pulses at 121.57 nm and 365.90 nm via a two step excitation. The neutral Rydberg atoms travel 250 mm before they are field-ionized and detected by a multi-channel plate detector. The arrival time is recorded by a multi-channel scalar. The rotatable detector allows data to be recorded at various scattering angles $\vartheta_s$. The graphene sample is epitaxially grown \textit{in situ} on a clean Pt(111) substrate by dosing ethylene (partial pressure $3 \times 10^{-8}$ mbar) at 700 $^{\circ}$C for 15 mins \cite{jiang2019imaging}.

\textit{Computational methods. --}
The all-atom potential energy surface (PES) employed in this study is obtained from the \textit{``}Geometry, Frequency, Noncovalent, eXtended Tight Binding" (GFN-xTB) method \cite{grimme2017robust}. While an accurate C-H adsorption barrier and binding well can be obtained with a hybrid density functional theory (DFT) \cite{becke1996density, lee1988development}, the associated computational costs make it impossible to simulate scattering energy and angular distributions. We found however, that a reparameterization of GFN-xTB within the {\it entos} software package \cite{manby2019entos} to best reproduce the minimum energy path of a hybrid functional DFT calculation resulted in an accurate all-atom PES at low computational cost. Details are presented in the SI Sec.~A. 
Notably, the GFN-xTB PES is 1000-fold more efficient to compute than that with a hybrid DFT functional. 
We are thus able to run far more trajectories than would otherwise be possible, making the reported calculations tractable.
 
We use \textit{ab initio} ring-polymer molecular dynamics (RPMD) under the influence of GFN-xTB potential energies to model the real-time quantum dynamics of the system.
RPMD \cite{craig2004quantum, habershon2013ring} is a trajectory-based dynamics method, in which the NQEs are taken into account based on Feynman's imaginary time path-integral formalism \cite{feynman1965quantum}. 
Although approximate, the method successfully describes both zero-point energy and tunneling effects in simulations at thermal equilibrium \cite{habershon2013ring, suleimanov2016chemical, markland2018nuclear} and more recently, it was applied to systems with non-equilibrium initial conditions \cite{welsch2016non} and in the microcanonical ensemble \cite{tao2020microcanonical}.

RPMD allows for the inclusion of NQEs by propagating classical trajectories of an  isomorphic system. 
The isomorphic system consists of $n$ replicas of the physical one, and is constructed such that exact quantum Boltzmann statistics are preserved \cite{feynman1965quantum, chandler1981exploiting, parrinello1984study}.

If  
${\bm q} = \left({\bm r}_{\text H1}, {\bm r}_{\text C1}, {\bm r}_{\text C2}, ... \right)$ 
denotes the column vector of positions of all atoms and $V({\bm q})$ the GFN-xTB potential energy for a geometry ${\bm q}$, then the RPMD equations of motion (EOM) are 
\begin{align} \label{rpmdeom}
\ddot{\bm q}_\alpha= 
\omega_n^2 \left( {\bm q}_{\alpha-1} +  {\bm q}_{\alpha+1} - 2  {\bm q}_{\alpha} \right) 
- {\bm m}^{-1} \hspace{-2pt} \cdot \nabla_{\hspace{-2pt} {\bm q_\alpha}} V({\bm q_\alpha}), 
\end{align}
where $\alpha=1, 2, ..., n$ is the index for different replicas, $\omega_n = n k_{\rm B} T$ is the strength of the harmonic springs that connect neighbouring replicas with $T$, being the system temperature, and 
${\bm m} = {\rm diag} \left( m_{\rm H1}, m_{\rm C1}, m_{\rm C2}, ... \right)$ the mass matrix for all the atoms involved. Note that with Eq.~\ref{rpmdeom}, all the atoms in the system are described quantum mechanically on the same footing.

Not only does RPMD provide an accurate and efficient way to perform quantum simulations, the fact that it captures NQEs \textit{via} trajectory propagation in classical phase space bridges the gap between classical and quantum mechanics. RPMD recovers exact quantum statistics in the limit $n \to \infty$ \cite{feynman1965quantum} and reduces to ordinary classical molecular dynamics when $n=1$. Hence, RPMD can be used for mixed quantum/classical (MQC) calculations where some degrees of freedom are described quantum mechanically and others classically  \cite{collepardo2008proton, suleimanov2012surface, miller2008isomorphic, menzeleev2011direct, kretchmer2013direct}.

For a system partitioned into quantum ${\bm q}$ and classical ${\bm Q}$ parts, 
 MQC-RPMD evolves the dynamics for the quantum mechanical portion using RPMD EOM (for $n$ replicas of the original system), and evolves the dynamics for the classical portion with Newton's EOM, i.e. \cite{miller2008isomorphic} 
\begin{align} \label{mqcrpmdeom}
\ddot{\bm q}_\alpha&= 
\omega_n^2 \left( {\bm q}_{\alpha-1} +  {\bm q}_{\alpha+1} - 2  {\bm q}_{\alpha} \right) 
- {\bm m}^{-1} \hspace{-2pt} \cdot \nabla_{\hspace{-2pt} {\bm q_\alpha}} V({\bm q}_{\alpha}, {\bm Q}), \nonumber \\
\ddot{\bm Q} &= 
- {\bm M}^{-1} \hspace{-2pt} \cdot \nabla_{ \hspace{-2pt} {\bm Q} }
 V( \bar{\bm {q}}, {\bm Q}), \qquad
\bar{\bm {q}} = \frac1n \sum_{\alpha=1}^n {\bm {q}}_\alpha.
\end{align}
where ${\bm m}$ and ${\bm M}$ are the mass matrices for the quantum and classical degrees of freedom, respectively.

As in Ref.~\onlinecite{jiang2019imaging}, we perform quantum dynamics simulations by a non-equilibrium RPMD approach \cite{welsch2016non}.
The graphene surface is modelled with a free-standing cluster of 42 carbon atoms. This model is sufficiently large to describe the PDOS spectrum of graphene obtained from calculations using periodic boundary conditions\textit{---}Fig.~S4. The boundary of the carbon cluster is terminated with H atoms, which are held fixed throughout the calculation using a RATTLE scheme \cite{andersen1983rattle}. A suspended graphene flake with 80 C\textit{--}atoms is used to eliminate the edge effects under conditions with larger values of $\vartheta_i$.

RPMD simulations are initialized by separately preparing the initial configurations for the non-interacting graphene sheet and H/D atom. 
The initial flake geometries are sampled from a thermalized ring-polymer trajectory at 300 K, performed using the Andersen thermostat
\cite{andersen1980molecular}.
The position and velocity of 
the centroid of the H/D atom ring-polymer is then determined according to the values of scattering energy and incidence angle in the experiment.
The internal modes for the H/D-atom ring-polymer are thermalized at the surface temperature 300 K following the non-equilibrium RPMD formulation \cite{welsch2016non,jiang2019imaging}, and it is confirmed that the results are insensitive to this choice of internal temperature\textit{---}see SI Fig.~S7.
The temporal evolution of the system is then found from either Eq.~\ref{rpmdeom} or \ref{mqcrpmdeom} using a time step of 0.5 fs \cite{korol2020cayley}. Propagation continues until the fate of the H or D atom is decided by trajectory analysis. For scattered atoms, the energy loss and outgoing angle is recorded for each trajectory. 
Convergence of the path-integral discretization was confirmed to be reached with 8 beads.
All reported simulations were performed using the {\it entos} software package \cite{manby2019entos}. Further details of the trajectory calculations are reported in the SI Sec.~B.

\begin{figure}[t]
\includegraphics[width=8.6cm]{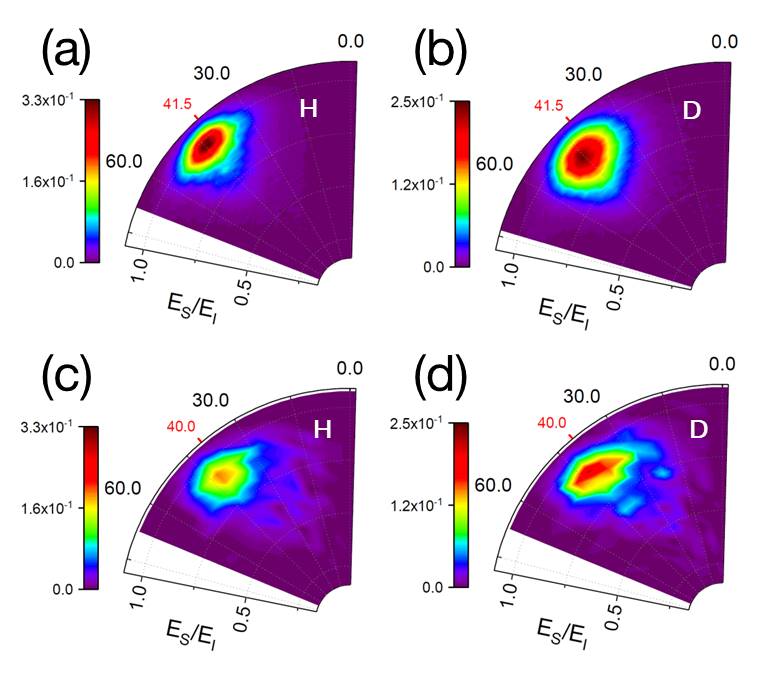}
\caption{\label{fig1} 
{\textbf{Comparing theory with experiment for H and D scattering distributions from graphene.} 
(\textbf{a}-\textbf{b}) Experimental distributions.  (\textbf{c}-\textbf{d}) Theoretical distributions. Result shown are from RPMD trajectories under the influence of GFN-xTB potential energies. The incidence energy of H/D translation, E$_i$, was $\sim 1$ eV. In all images, the scattering energy, E$_s$, is shown along the radial coordinate and the scattering angle, $\vartheta_s$, is shown on the polar coordinate. The distributions are normalized to their integrals. The red ticks indicate the specular scattering angle.}}
\end{figure}

\textit{Results. --}
Fig.~\ref{fig1}(a) and (b) show examples of experimental scattering distributions for H and D, respectively, colliding with graphene. Both scattering distributions peak close to the specular angle and energy loss is small but slightly larger for D than for H. Additional scattering distributions for other values of $\vartheta_i$ are shown in Fig.~S2.
The scattered flux comes from atoms reflected at the barrier to chemisorption; it therefore decreases as $\vartheta_i$ is reduced, because ever more atoms pass over the barrier to chemisorption, with insufficient energy to return \cite{jiang2019imaging}.

Fig.~\ref{fig1}(c) and (d) show RPMD simulations of the experimental results of 
Fig.~\ref{fig1} (a) and (b). Also see Fig.~S2. Agreement is excellent. 
These trajectory calculations also show that the scattered flux is due to H/D atoms that never reach the chemisorption well\textit{---} which remain trapped.  
Thus, the observed signals are a direct measure of the H/D survival probability. 

\begin{figure}[t]
\includegraphics[width=8.6cm]{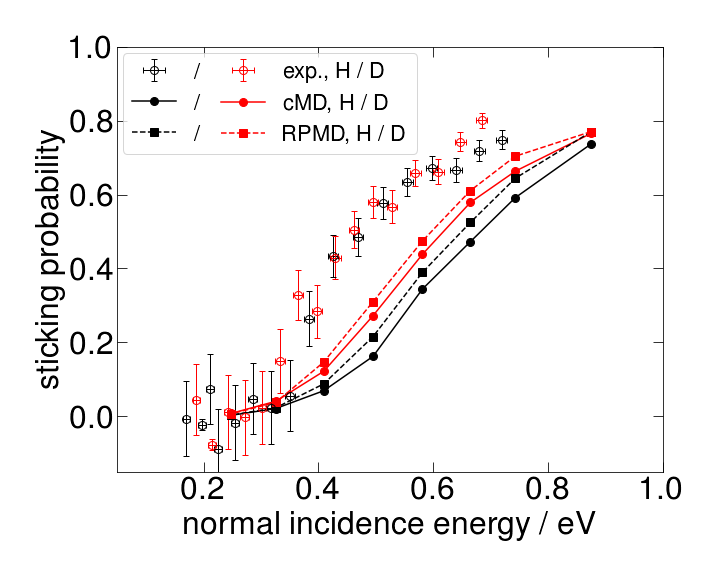}
\caption{\textbf{Comparison of experiment and theory for H/D sticking probabilities to graphene.} Experimental sticking probabilities (circles with error bars) for H (black) and D (red) are compared to classical (solid lines and symbols) and ring polymer (dashed lines and symbols) molecular dynamics simulations. The incidence energy was held constant and the incidence angle was varied to control the normal component of incidence energy. The statistical error associated with the trajectory calculations is less than $1.1$\%.}
\label{fig2}
\end{figure}

Experimentally derived sticking probabilities are shown in Fig. 2. See SI Sec.~C and Fig.~S3. The sticking probability increases with the normal component of the incidence energy.
Thresholds near $\sim$0.35 eV are seen for both H and D, a clear sign of the adsorption barrier. No H/D isotope effect can be discerned. Fig. 2 also shows sticking probabilities from both classical molecular dynamics (cMD) and RPMD simulations\textit{---}both are in good agreement with experiment.
A small inverse H/D isotope effect is clearly present\textit{---}D sticking is $\sim$10\% more likely than H sticking. 

\textit{Discussion. --} 
By comparing the predictions of sticking probabilities from different computational approaches, we gain insight into how classical and quantum effects manifest in the dynamics of  H and D sticking to graphene. cMD simulations predict an increased sticking for D than H atom (Fig. 2: solid red \textit{vs.} solid black lines). 
This results from the  $\sim$1.4x longer interaction time of the D atom compared to the H atom\textit{---}See Fig.~S6\textit{---}a result of classical inertia \cite{kelly2009newtonian, andujar2012racing}. 
The longer interaction time of D  allows greater relaxation of the C flake during the trajectory, reducing the height of the effective barrier to sticking. 

Fig. 2 also shows that NQEs are present for both H and D scattering\textit{---}note the deviation of RPMD from cMD predictions for both isotopes. 
\begin{figure}[!t]
\includegraphics[width=8.6cm]{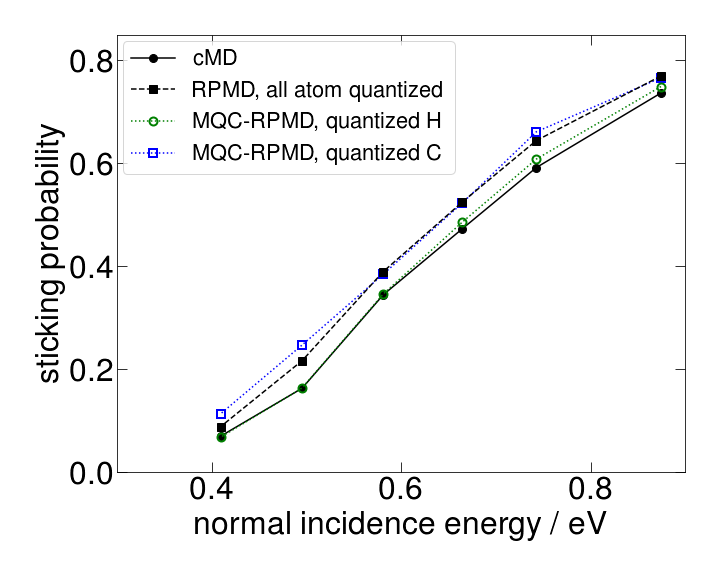}
\caption{\textbf{Simulated H-on-graphene sticking probability as a function of normal incident energy}. Results are obtained with cMD, all-atom quantized RPMD (with Eq.~\ref{rpmdeom}), and the mixed quantum/classical implementations of RPMD (with Eq.~\ref{mqcrpmdeom}). }
\label{fig3}
\end{figure}
 To investigate the mechanistic origin of these NQEs, we also calculated sticking probabilities with the MQC implementation of RPMD 
\cite{collepardo2008proton, suleimanov2012surface, miller2008isomorphic, menzeleev2011direct, kretchmer2013direct}.
See Fig.~\ref{fig3}.
 In the MQC-RPMD simulations, either the projectile atom  or the graphene flake was described quantum mechanically, while the remainder of the system was described classically. When the projectile is quantized but the graphene flake moves classically (dashed green line), no difference is found in comparison to the conventional classical MD result (black solid line). However, when the C atom motion is quantized but the projectile atom is treated classically (dashed blue line), no difference to the full RPMD result can be seen (dashed black line). 
This shows that the  NQEs 
 that most influence the sticking process
 in this system result from the C-atom motion and that the quantum mechanical motion of the C-atoms enhances both H and D sticking.
We note that a previous study that performed such a breakdown for H-on-surface diffusive dynamics under equilibrium conditions reached the opposite conclusion about the relative importance of H versus surface-atom quantization \cite{suleimanov2012surface}, which is interesting but not inconsistent with our findings due to the different mechanism that governs the non-equilibrium collision process studied here.
 
While it must be acknowledged that RPMD is an approximate description of the dynamical processes considered here, there is reason to expect that the method is being applied within a regime of good confidence. Firstly, for both equilibrium and non-equilibrium processes, RPMD is exact in the limit of short-time dynamics, where the relevant timescale is $\beta\hbar\sim 25$ fs at room temperature; given that the entire timescale of H/D-surface contact that dictates whether sticking occurs falls within a window of ~10 fs (Fig.~4A of Ref.~\onlinecite{jiang2019imaging}), the current application appears to be safely within that regime.  Secondly, RPMD is exact in the classical mechanical limit since the description reduces to Newton's equations of motion; again, given that the NQEs in the current problem are quite modest, it suggests that the current application is within this regime.  
And finally, although the exactness of RPMD for both the short-time and classical limits holds regardless of whether the simulated property corresponds to a correlation function of linear or non-linear operators, we note that the RPMD approximation is typically more accurate for time correlation functions (TCFs) of linear operators than for non-linear operators; \cite{habershon2013ring} while this might raise concern because the fluctuation-dissipation description of vibrational energy relaxation involves a TCF of non-linear operators \cite{tanimura2009modeling}, the centroid position of H/D atom that dictates whether sticking occurs is in fact a linear function of position. 
Taken together, these considerations suggest that RPMD provides a reliable description of the physical processes considered here, while allowing for the full-dimensional simulations that are needed to capture key aspects of the sticking mechanism  \cite{jiang2019imaging}.

Regardless of these methodological considerations, we emphasize that the mechanistic picture that emerges is physically intuitive and consistent with our previous interpretation \cite{jiang2019imaging}.
Although perhaps at first surprising, the results are easily understood when considering the dynamical mechanism for sticking. 
Ultrafast energy loss of H atom translation is associated with in-plane C-atom excitation adjacent to the reaction center. 
When the  in-plane C-atom motion important to the sticking is more realistically modelled using quantum simulations rather than classical simulations, sticking is enhanced and better agreement to experiment is achieved. 

The importance of NQEs originating from C atoms emphasizes the breakdown of the classical approximation for describing processes that are sensitive to high frequency phonons. When we use the same approach with a PES that fails to capture the quantized frequency distribution of the graphene flake, these NQEs do not appear.  See Figs.~S4 and ~S5.   
We also note that the conclusion that C-atom quantization is the leading source of NQEs in the sticking process is consistent with earlier work \cite{bonfanti2018sticking} using reduced-dimensionality models, which otherwise fail to capture a quantitative description of the sticking process. 

We began this study with the intention to investigate H/D isotope effects in adsorption at graphene; however, it turns out the the quantum motion of the heavy atoms is more important. This reflects the polaytomic motion involved in the formation of a C-H chemical bond in this adsorption process. Such behavior is unlikely to be unique to H sticking to graphene.

X.T. acknowledges support from the Department of Dynamics at Surfaces at the MPI for Biophysical Chemistry and ICASEC at University of Goettingen during the visit. HJ, OB and AMW acknowledge support the from the SFB1073 under project A04, from the Deutsche Forschungsgemeinschaft (DFG) and financial support from the Ministerium für Wissenschaft und Kultur (MWK) Niedersachsen, and the Volkswagenstiftung under Grant No. INST 186/902-1 to build the experimental apparatus. AMW, MK and AK also acknowledge the Max Planck Society for the Advancement of Science.
F.D. and T.F.M. acknowledge that this material is based
on work performed by the Joint Center for Artificial Photosynthesis,
a U.S. Department of Energy (DOE) Energy Innovation Hub,
supported through the Office of Science of the DOE under award
DE-SC0004993; and X.T. and T.F.M. acknowledge support from the DOE (award DE-SC0019390).
We thank Dan Auerbach and Dirk Schwarzer for helpful discussions.

H.J. and X.T. contributed equally to this work. Lists of authors to whom correspondence should be addressed: alec.wodtke@mpibpc.mpg.de (A.M.W.); akandra@gwdg.de (A.K.); tfm@caltech.edu (T.F.M.); and oliver.buenermann@chemie.uni-goettingen.de (O.B.).


%

 \end{document}